\documentclass[a4paper,11pt]{article}
\usepackage{a4wide}
\usepackage{amsmath}
\usepackage{graphicx}
\title{Deriving the Regge-Wheeler and Zerilli equations\\in the
general static spherically-symmetric case\\with
Mathematica\texttrademark \;and MathTensor\texttrademark}
\author{Gianluca Cruciani\\ \textit{Universit\`a "La Sapienza", Roma}}
\begin{document}
\maketitle
\begin{abstract}
An efficient approach to tensor perturbation calculations by
proper use of computer algebra methods is described,
 reaching the sufficient generality required for a comprehensive analysis of
the Schwarzschild and Reissner-Nordstr\o m metric stability
problem.
\end{abstract}
\section{Introduction}
The stability problem for a Schwarzschild black hole in the form
of a ``pure metric" perturbation analysis was settled by T. Regge
and J. A. Wheeler in a classic 1957 article  \cite{rw}. The main
practical achievement of this work was undoubtedly the formulation
of a gauge transformation approach that allows a complete
radial/angular separation of the Einstein equations in the two
cases of odd and even parity, preliminarily established, reaching
a formal solution in the axial (or \emph{magnetic}) case: the
so-called \emph{Regge-Wheeler Equation}. Due to mathematical
complications, however, the full analysis was only completed
thirteen years later, after the work of Mathews \cite{mathews},
Edelstein-Vishveshwara \cite{edelvish} and Zerilli \cite{zerilli3}
that either provided a more rigorous approach to the use of tensor
harmonics, or resolved some compatibility problems in the analytic
treatment of the system, or else provided the final form of the
equation for the radial perturbation functions in the polar (or
\emph{electric}) case: the so-called \emph{Zerilli Equation}.  The
search for simplicity also led these authors to exploit some
useful but not general relations between curvature tensors, like
those derived by Eisenhart \cite{eisen}, valid to the first order
and/or only in the Schwarzschild case (latin tensor indices are
used for consistency with the implemented algorithms):
\begin{align}
&\delta G_{mn}=\delta R_{mn}\\&\delta R_{mn}= \delta \Gamma
^{p}_{mn;p}-\delta \Gamma ^{p}_{mp;n}\\&\delta \Gamma ^{i}_{jk}=
\frac{1}{2}g^{ip}(h_{jp;k}+h_{kp;j}-h_{jk;p})
\end{align}
where $h$ is the perturbation tensor and $\delta \Gamma,\;\delta
R,\;\delta G$ are the perturbed parts of the affine connections,
Ricci and Einstein tensors.\\ On this way, after having performed
the gauge transformations for each parity case, we are left with
two systems of, respectively, three and six independent radial
ordinary differential equations in two (namely $h_0(r),\;h_1(r)$)
and three (namely $H(r),\;H_1(r),\;K(r)$) arbitrary perturbation
functions, to be determined. The magnetic system is
straightforwardly reduced to a single first-order equation in
$h_1(r)$ which, by a simple variable substitution, leads to the
final result:
\begin{equation}\label{schr}
Q''(r^*)+[k^2-V(r)]Q(r)=0
\end{equation}
(\emph{Regge-Wheeler Equation}), where $r^*$ and $Q$ (the first
called the \emph{tortoise coordinate} by Wheeler as a citation of
the Zeno paradox) are (implicitly or not) defined by:
\begin{equation}\label{varch}
d/dr^*=a(r)d/dr \mbox{\hspace{1cm}} Q(r)=b(r)h_1(r)
\end{equation}
where $a(r),\ b(r)$ are arbitrary functions to be determined in
each particular case and $V(r)$, in this Schr\"{o}dinger-like
equation, plays the role of an effective potential.\\ Finally, the
electric system, by a more complex change of variables procedure,
necessary to deal with the terms in the wave number $k$, was found
by Zerilli \cite{zerilli3} to be represented by an equation
formally equal to (\ref{schr}) (the \emph{Zerilli Equation}) only
with a different (but still algebraic) expression of the
potential. Since all the variables of the system are mutually
expressed by regular algebraic relations, the stability problem,
analyzed by substituting different forms of $k$ into (\ref{schr}),
can be extended in its validity to the whole perturbation. The aim
of the present work is to show that, with the essential help of
computer algebra software, a similar analysis can be carried out
for both parity cases, dealing with the more general spacetime of
a spherical, non-rotating, eventually charged collapsed object,
therefore allowing specialization not only to the Schwarzschild
but also to the Reissner-Nordstr\o m metric.
\section{The static spherically-symmetric system}
To begin a less rigid analysis than that induced by the
formulation of equations (1--3), we deal first of all with the
full expression of the Einstein tensor which, viewed as a function
of the metric and its ordinary partial derivatives, reads as:\\
\includegraphics[width=15cm]{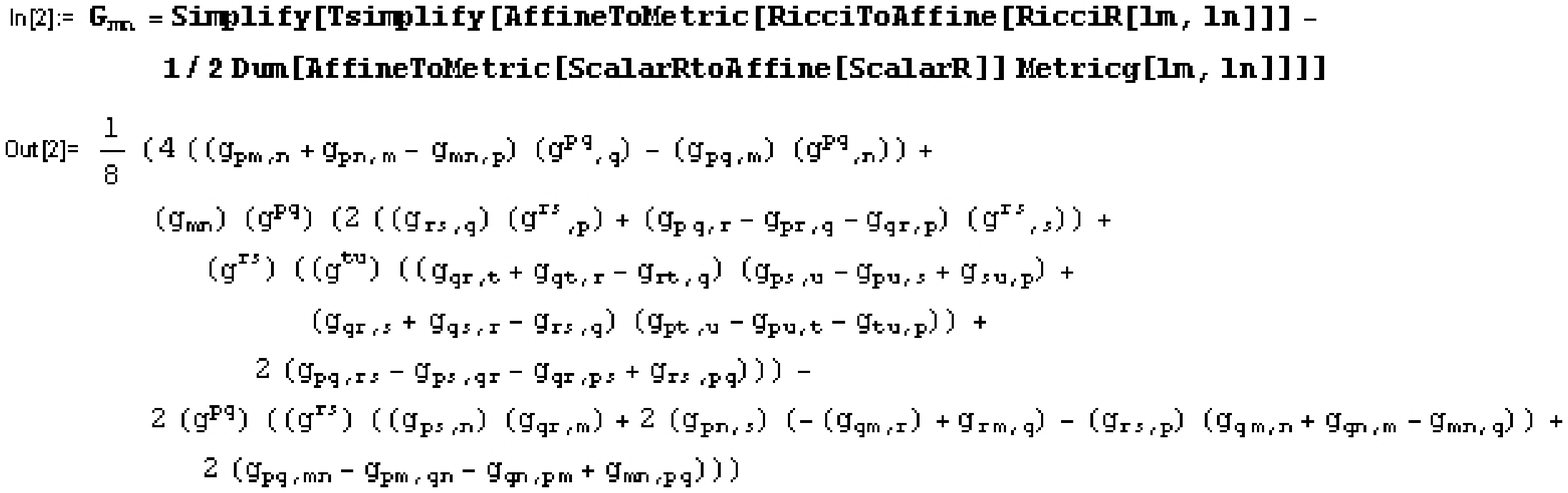}\\
Here, to the unperturbed metric, which refers to the usual
covariant expression of the line element belonging to a generic
spherically-symmetric expression depending on the radial arbitrary
functions $\lambda(r),\ \nu(r)$ :
\begin{equation*}
ds^2=e^{\lambda(r)}dr^2+r^2(d\theta^2+\sin^2\theta\;d\phi^2)-e^{\nu(r)}dt^2
\end{equation*}
must be added the perturbation terms, whose time dependency is
represented only by a $e^{-\imath kt}$ factor while the angular
one is limited to functions of the polar angle $\theta$.\\ Since
"\textbf{Components}", MathTensor's routine for calculating
curvature tensors from an input metric, has the useful feature of
applying arbitrary combinations of Mathematica commands to each
component \cite{mathtens}, the first order form of the Einstein
tensor can be readily obtained by appending to the metric input
file, where the Regge-Wheeler-gauge perturbation terms are added
with a ``small parameter" $q$ as a factor, a line like:
\begin{verbatim}
CompSimp[a_]:=Simplify[Normal[Series[Expand[a],{q,0,1}]]/.q->1]
\end{verbatim}
\subsection{The magnetic case}
There are only three non-zero components
$\left\{(r,\;\phi),\;(\theta,\;\phi),\;(\phi,\;t)\right\}$ of the
tensor equation that replaces the (1):
\begin{equation}\label{sist}
\delta G_{mn}=G_{mn}^{q=1}-G_{mn}^{q=0}=0
\end{equation}
(where the subtracted quantity is the unperturbed metric tensor)
and that substantially reproduce, in their radial form, the
solution system \cite{edelvish}-(sys. 2), once the angular terms
coming from the perturbation, where they are represented by
$f(\theta)=\sin\theta\;\partial P_{L}(\cos\theta)/\partial\theta$
($P_{L}$ being the Legendre polynomial to the multipolar order
$L$), are fully simplified through the following relations:
\begin{align*}
f(\theta)&=L\,[-P_{L-1}(\cos\theta)+L \cos\theta\;
P_{L}(\cos\theta)]\\
f'(\theta)&=-L\,(L+1)\sin\theta\;P_{L}(\cos\theta)\\
f''(\theta)&=L\,(L+1)\,[L\;P_{L-1}(\cos\theta)-(L+1)\cos\theta\;
P_{L}(\cos\theta)]
\end{align*}
and, being completely factored, are consequently eliminated.\\ The
first two equations (from now on, almost everywhere in the rest of
this paper, the notation $\lambda=(L-1)(L+2)/2$ will be adopted
-not to be confused with the definition of the metric's radial
function $\lambda(r)$!) turn out to be of the first order with
non-mixed dependence on the functions' derivatives in such a way
that a single second-order differential equation can be obtained
by:\\
\includegraphics[width=15cm]{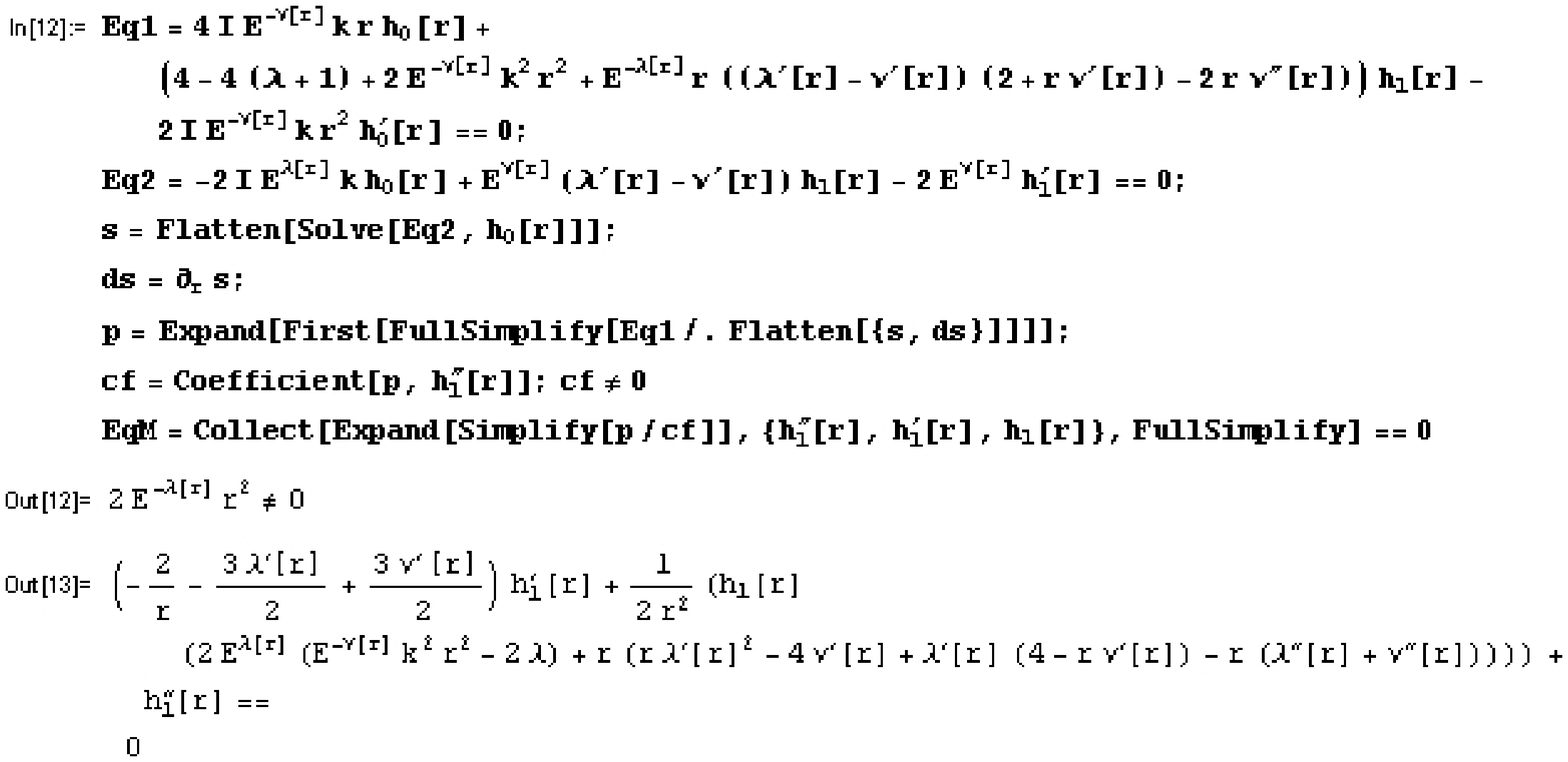}\\
Analyzing the dependence on $k$, it is readily seen that to impose
the equation (\ref{schr}) with the conditions (\ref{varch}), a
procedure of polynomial coefficient identification is required,
carried out in the following way:\\
\includegraphics[width=15cm]{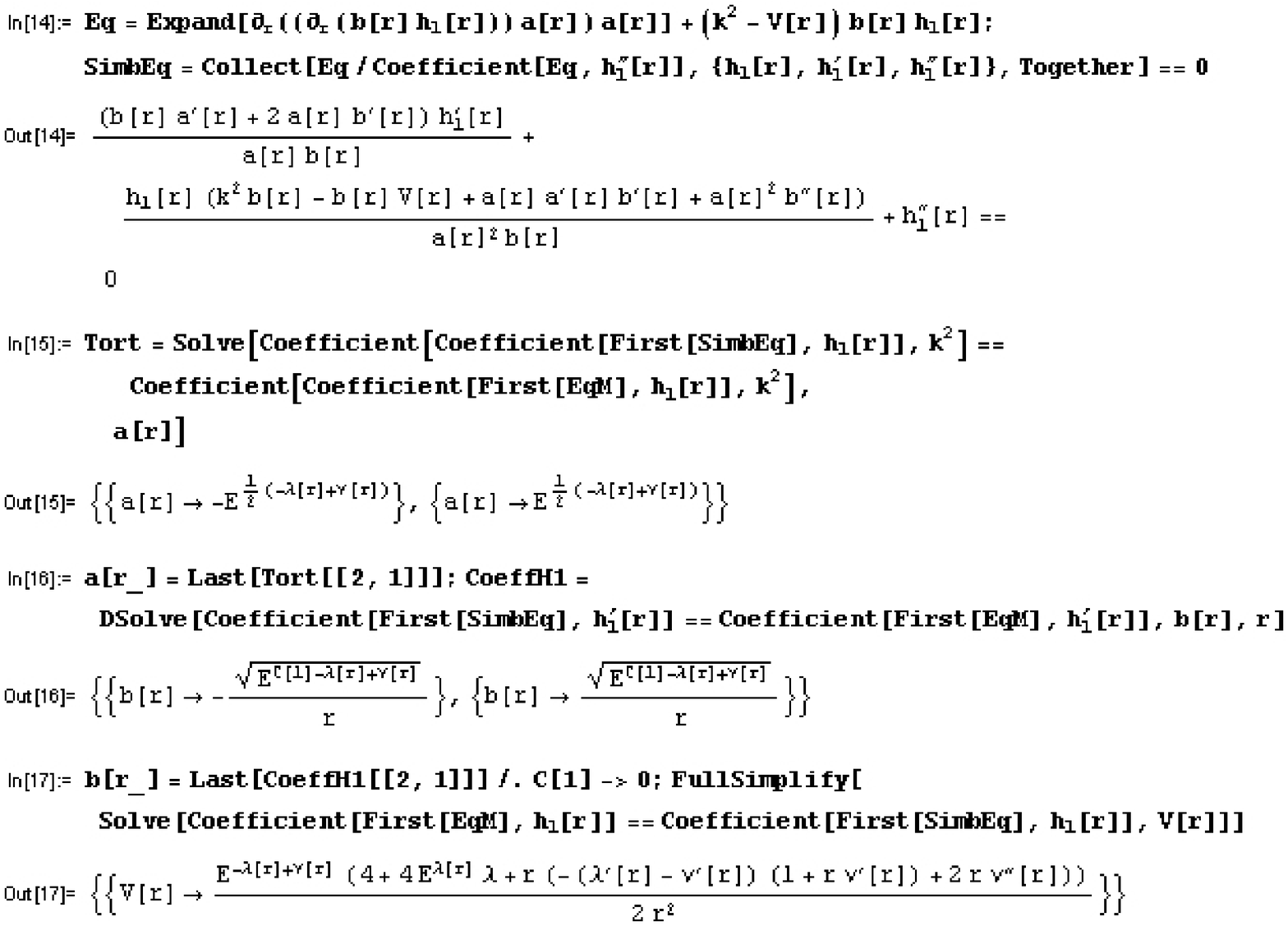}\\
\subsection{The electric case}
Here we deal with the seven non-zero components of the perturbed
Einstein tensor (the complementary set with respect to the
previous three ``magnetic" components), whose angular parts,
coming from the perturbing gauge function
$F(\theta)=P_L(\cos\theta)$, are factored, once the substitutions
\begin{align*}
F'(\theta)&=\frac{L}{\sin\theta}\;[-P_{L-1}(\cos\theta)+\cos\theta\;P_{L}(\cos\theta)]\\
F''(\theta)&=\frac{L\cos\theta}{\sin^2\theta}\;P_{L-1}(\cos\theta)-L\,(\cot^2\theta+L+1)\;P_{L}(\cos\theta)
\end{align*}
are performed, the resulting expressions simplified and the
original four radial perturbation functions are reduced to three
by the identification $H_0(r)=H_2(r)\equiv H(r)$.\\ Of the six
independent linear differential equations so obtained (the
diagonal $\theta$ and $\phi$ terms being equal), four turn out to
be of the first order, unlike the Schwarzschild-specialized system
which instead has three first order equations, and a completely
algebraic variable-elimination procedure allows the derivation of
a first integral condition which is the generalization of
\cite{edelvish}-(eq. 10). As in Zerilli's procedure
\cite{fackerell}, more elaborate but always tied to a polynomial
coefficient identification principle, this algebraic condition
plus three of the previous equations can be treated, with a double
function substitution and the change of the radial coordinate, to
form a new system of seven differential equations in four
variables (plus the derivative of the new radial coordinate with
respect to $r$), this time non-linear but independent of $k$, to
which the formal definition of the effective potential must be
added. Three of these unknown functions are then found, after a
cascade of algebraic eliminations (which don't show the residual
arbitrariness found by Zerilli in the Schwarzschild treatment), to
be quite simply dependent on the fourth, which satisfies a final
second-order very complex equation, fortunately analytically
solvable when specialized to the Schwarzschild and
Reissner-Nordstr\o m metrics.
\section{Results of the analysis}
A comprehensive table of functions and variables in the three
cases -- general static spherically-symmetric form (G),
Schwarzschild (S) and Reissner-Nordstr\o m (RN)-- can be sketched,
the latter two obtained, respectively, by the two substitutions
\{$\lambda(r)=-\ln\,(1-2m/r)\;;\;\nu(r)=-\lambda(r)$\} and
\{$\lambda(r)=-\ln\,( 1-2m/r+Q^2/r^2)\;;\;\nu(r)=-\lambda(r)$\}\;:
\vspace{5mm}\\
\renewcommand{\arraystretch}{2.5}
\begin{tabular}{llll}
& \emph{$\pi$} & \emph{Schr\"{o}dinger wave-like function} &
\emph{Tortoise coordinate}\\ \hline G & mag. &
$Q(r)=\exp\left[\frac{1}{2}(\nu (r)-\lambda
(r))\right]\frac{h_{1}(r)}{r}$ & $r^*=\int \exp\left[\frac{1}{2}
(\lambda (r)-\nu (r))\right] dr$ \\ \cline{2-4} & el. &
$\hat{K}(K(r),\frac{H_{1}(r)}{r})$ & $r^*={\displaystyle \int}
\!e^{\frac{\lambda(r)}{2}}\!\left[\frac{2\lambda
e^{\lambda(r)}-r\lambda '(r)+2r\nu '(r)}{e^{\nu(r)}\left(2\lambda
e^{\lambda(r)}+3r\nu '(r)\right)}\right]^{\frac{1}{2}}\!\!dr$
\\ \hline S & mag. & $Q(r)=\left(1-\frac{2m}{r}\right)\frac{h_{1}(r)}{r}$ & $r^*=r+2m\ln(r-2m)$ \\ \cline{2-3} &
el. & $\hat{K}(K(r),\frac{H_{1}(r)}{r})$ & \\ \hline RN & mag. &
$Q(r)=\left(1-\frac{2m}{r}+\frac{Q^2}{r^2}\right)\frac{h_{1}(r)}{r}$
& $r^*=r+m\ln(r^2-2mr+Q^2)+\eta(r)$
\\ \cline{2-3} & el. & $\hat{K}(K(r),\frac{H_{1}(r)}{r})$ & $\left[\eta(r)=\frac{2m^2-Q^2}{\sqrt{Q^2-m^2}}\arctan\left(
\frac{r-m}{\sqrt{Q^2-m^2}}\right)\right]$
\\ \hline
\end{tabular}
\vspace{1cm}\\ and the correspondent couples of specialized
expressions of the Regge-Wheeler and Zerilli's potentials are:\\
\begin{itemize}
\item Schwarzschild:
\begin{align*}
& V_{mag}^S(r)=2\left( 1-\frac{2m}{r}\right) \left(
\frac{\lambda+1}{r^2}-\frac{3m}{r^3}\right)\\ &
V_{el}^S(r)=2\left( 1-\frac{2m}{r}\right) \frac{\lambda
^2(\lambda+1)r^3+3\lambda ^2mr^2+9\lambda m^2r+9m^3}{(\lambda
r+3m)^2r^3}
\end{align*}
\item Reissner-Nordstr\o m:
\begin{align*}
& V_{mag}^{RN}(r)=2\left( 1-\frac{2m}{r}+\frac{Q^2}{r^2}\right)
\left(
\frac{\lambda+1}{r^2}-\frac{3m}{r^3}+\frac{3Q^2}{r^4}\right)\\ &
V_{el}^{RN}(r)=\left(1-\frac{2m}{r}+\frac{Q^2}{r^2}\right)\frac{P(r)}{4\;[r(\lambda
r+3m)-3Q^2]^2\;r^4}
\end{align*}
with
\begin{equation*}
\begin{split}
P(r) & =8\lambda^2(\lambda+1)r^6+24\lambda^2mr^5+2\lambda\;
[36m^2-7(2\lambda-3)Q^2]\;r^4\\ &
+12m\;[6m^2-(19\lambda-3)Q^2]\;r^3-3Q^2\;[108m^2-(38\lambda-3)Q^2]\;r^2\\
& +342mQ^4r-117Q^6
\end{split}
\end{equation*}
\end{itemize}
As a final remark, it is straightforward to verify that the two
different kinds of tortoise coordinates in the general case reduce
to one for all metrics having $\lambda(r)=-\nu(r)$ and that as
expected, $ \lim_{Q \rightarrow 0} V^{RN}(r)=V^S(r)$ holds for
both the magnetic and the electric parities.

\end{document}